\documentclass[a4paper]{jpconf}
\usepackage{graphicx}
\begin{document}
\title{Dark Matter searches through cross-correlations of gamma rays with neutral-hydrogen intensity mapping }

\author{Elena Pinetti}
\address{University of Torino and INFN/Torino - Italy}
\ead{elena.pinetti@unito.it}

\begin{abstract}
In this work we derive the first theoretical prediction of the cross-correlation signal between the unresolved gamma-ray background and the 21-cm line originated by the spin-flip transition of neutral hydrogen atoms, by taking as benchmark experiment the space telescope {\it Fermi}-LAT for gamma rays and the next-generation radio telescope Square Kilometer Array (SKA) as well as its precursor MeerKAT for the 21-cm emission. The attainable bounds in the dark matter (DM) parameter space are envisioned to be competitive already with the combination {\it Fermi}-LAT $\times$ MeerKAT, but SKA will allow to go deeper and probe a thermal DM particle up to masses of 130 GeV. A future gamma-ray detector with better angular resolution and larger exposure together with an ungraded radio telescope will have the potentiality to probe a DM candidate with thermal annihilation cross-section and masses up to the TeV scale.  

\end{abstract}

\section{Introduction}
This work relates to the synergies across the electromagnetic spectrum for dark matter indirect detection searches and is based on Ref. \cite{Pinetti2019}.
In our Universe there are evidences for the presence of dark matter (DM) on different scales: in our Galaxy, in the external galaxies of the Milky Way, in clusters of galaxies as well as in the cosmic web. We still ignore the intimate nature of DM but one possibility is that it consists of a new elementary particle which does not interact with photons. If so, we expect that DM particles could annihilate and indirectly produce a huge variety of astrophysical messengers, among which hadrons and leptons which, in turn, can produce gamma rays as a decay product or as a final state radiation. This gamma emission would faintly contribute to the unresolved gamma-ray background (UGRB) together with the leading contribution of the unresolved astrophysical sources, notably star-forming galaxies and active galactic nuclei. As proposed in \cite{Camera2013}, the cross-correlation technique is an invaluable tool to scrutinize the UGRB since it attempts to correlate a gravitational tracer of the large-scale matter distribution in the Universe with the electromagnetic background radiation, potentially indirectly originated by annihilation events of DM particles. In Ref. \cite{Pinetti2019}, we delve into the opportunities offered by the cross-correlation between the UGRB and the 21-cm spectral line emitted by neutral hydrogen (HI) atoms in the spin-flip transition. In fact, since HI in the post-reionisation era is hosted in galaxies which, in turn, lie within DM haloes, it can be considered a suitable gravitational tracer of the underlying matter distribution in the Universe. The HI line can be studied with an up-and-coming technique known as intensity mapping \cite{Santos} which consists in mapping the intensity fluctuations of the HI brightness temperature. Moreover, the information on redshift is for free since the emitted frequency of the 21-cm line $\nu_e$ = 1420 MHz is directly related to the observed frequency $\nu_o$ through $\nu_o = \frac{\nu_e}{1+z}$.
In order to derive the forecasts for the bounds in the DM parameter space, we consider as benchmark experiments {\it Fermi}-LAT for gamma rays and the radio telescope SKA for HI intensity mapping measurements as well as its precursor MeerKAT. Lastly, we investigate the opportunities offered by a new-generation gamma-ray detector together with an ungraded radio telescope. 



\section{The cross-correlation signal} 
The function of the angular multipole $l$ to be taken into account in the analysis of the cross-correlation between two quantities $i$ and $j$ is the so-called angular power spectrum (APS):

\begin{equation}
 C_l^{ij} = \int \frac{\rm d \chi}{\chi^2} \, W_i (\chi) W_j(\chi) P_{ij} \left(k=\frac{l}{\chi}, \chi \right),
\label{eq:APS}
\end{equation}

In this context, the two observables are the 21-cm brightness temperature $T_b$ and the UGRB intensity $I_{\gamma}$. The objects that come into play in the APS are the radial comoving distance $\chi(z)$, the window function $W_i(\chi)$, which encloses the information on the redshift evolution of the observable under study, and the power spectrum $P_{ij} \left(k=\frac{l}{\chi}, \chi \right)$ in the Limber approximation. In the halo model framework, $P(k)$ can be split into one-halo and two-halo contributions which encompass information on the level of correlation on small and large-scales, respectively.
These two components in the case of the cross-correlation between the HI line and the DM gamma rays intensity take the form: 
\begin{equation}
    P^{\rm 1h}_{\rm HI-DM}(k) =  \int_{M_{\rm min}}^{M_{\rm max}} \rm d M \, \frac{\rm d N}{\rm d M} \, \frac{\tilde{v}_{\rm DM}(k)}{\Delta^2} \, \tilde{u}_{\rm HI}(k) \, \frac{M_{\rm HI}}{\overline{\rho}_{\rm HI}}
\end{equation}

\begin{equation}
    P^{\rm 2h}_{\rm HI-DM}(k) =  \left[\int_{M_{\rm min}}^{M_{\rm max}}  \rm d M \, \frac{\rm d N}{\rm d M} \, \frac{\tilde{v}_{\rm DM}(k)}{\Delta^2}  b_1 \right]\nonumber
    \left[\int_{M_{\rm min}}^{M_{\rm max}}  \rm d M \, \frac{\rm d N}{\rm d M} \, \tilde{u}_{\rm HI}(k) \frac{M_{\rm HI}}{\overline{\rho}_{\rm HI}} b_1 \right] P_{\rm lin}(k),
\end{equation}
where the integrals are performed over the DM halo mass $M$ and the dependence on $M$ and $z$ is implied in order to ease the notation. The quantities involved are the halo mass function $dN/dM$, the Fourier transform of the HI density distribution $\tilde{u}_{\rm HI}$ and the one of the matter distribution squared $\tilde{v}_{\rm DM}$ (since we are dealing with annihilation events), the linear bias $b_1$, the linear matter power spectrum $P_{\rm lin}$ and the clumping factor $\Delta^2$. $M_{\rm HI}$ quantifies the HI mass contained in a halo and $\overline{\rho}_{\rm HI}$ denotes the HI average density.
In the case of the astrophysical radiation, the luminosity $L$ takes over for the DM halo mass $M$ and the gamma-ray luminosity function $\rm d N/ \rm d L \equiv \phi(L,z)$ for the halo mass function $dN/dM$.
The classes of astrophysical sources under consideration are BL Lac, flat-spectrum radio quasar, misaligned active galactic nuclei and star-forming galaxies. For further details on the modelling, please refer to \cite{Pinetti2019}.

\section{Results}
\begin{figure*}[t]
\includegraphics[width=0.45\textwidth]{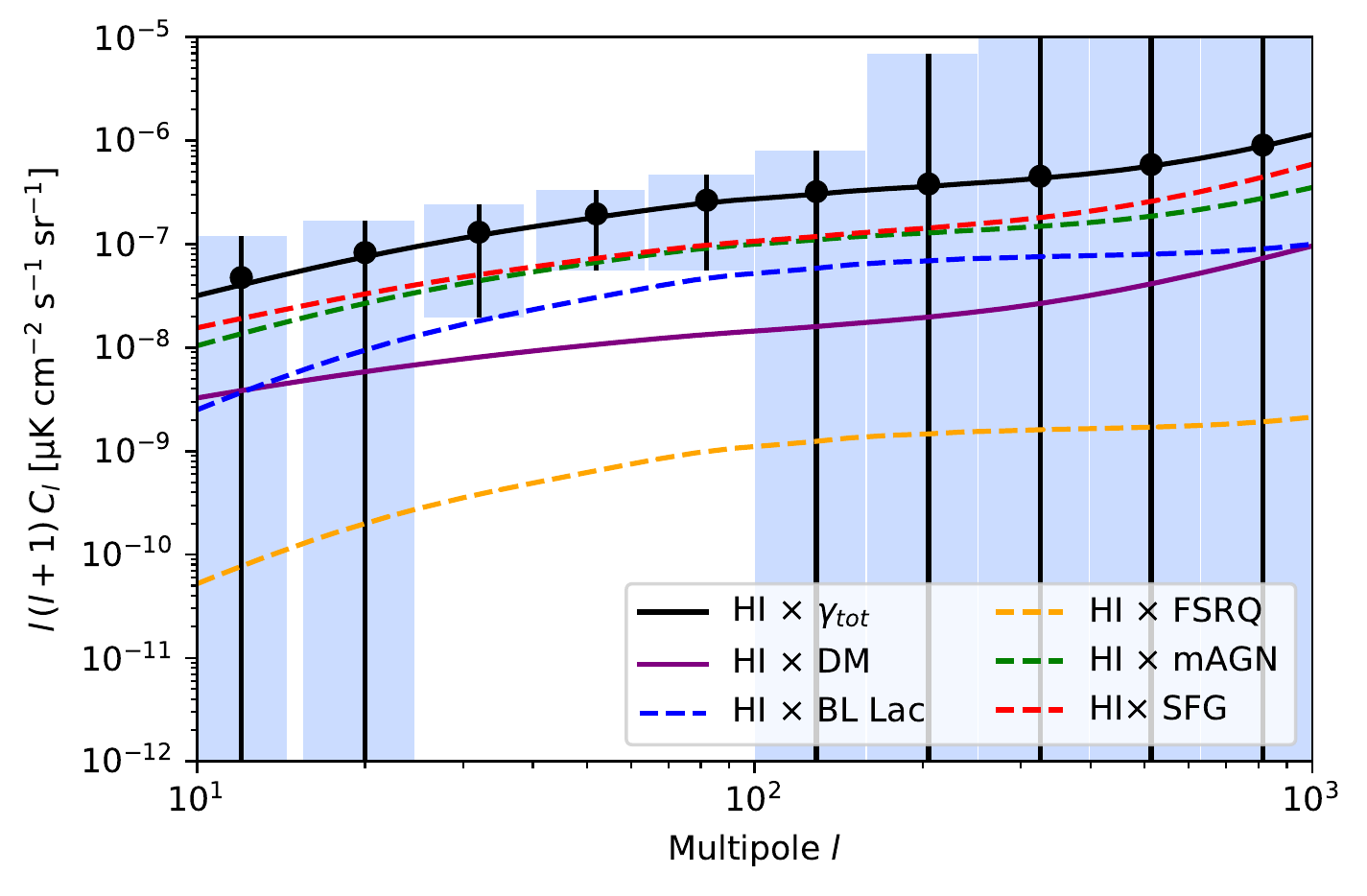}
\includegraphics[width=0.45\textwidth]{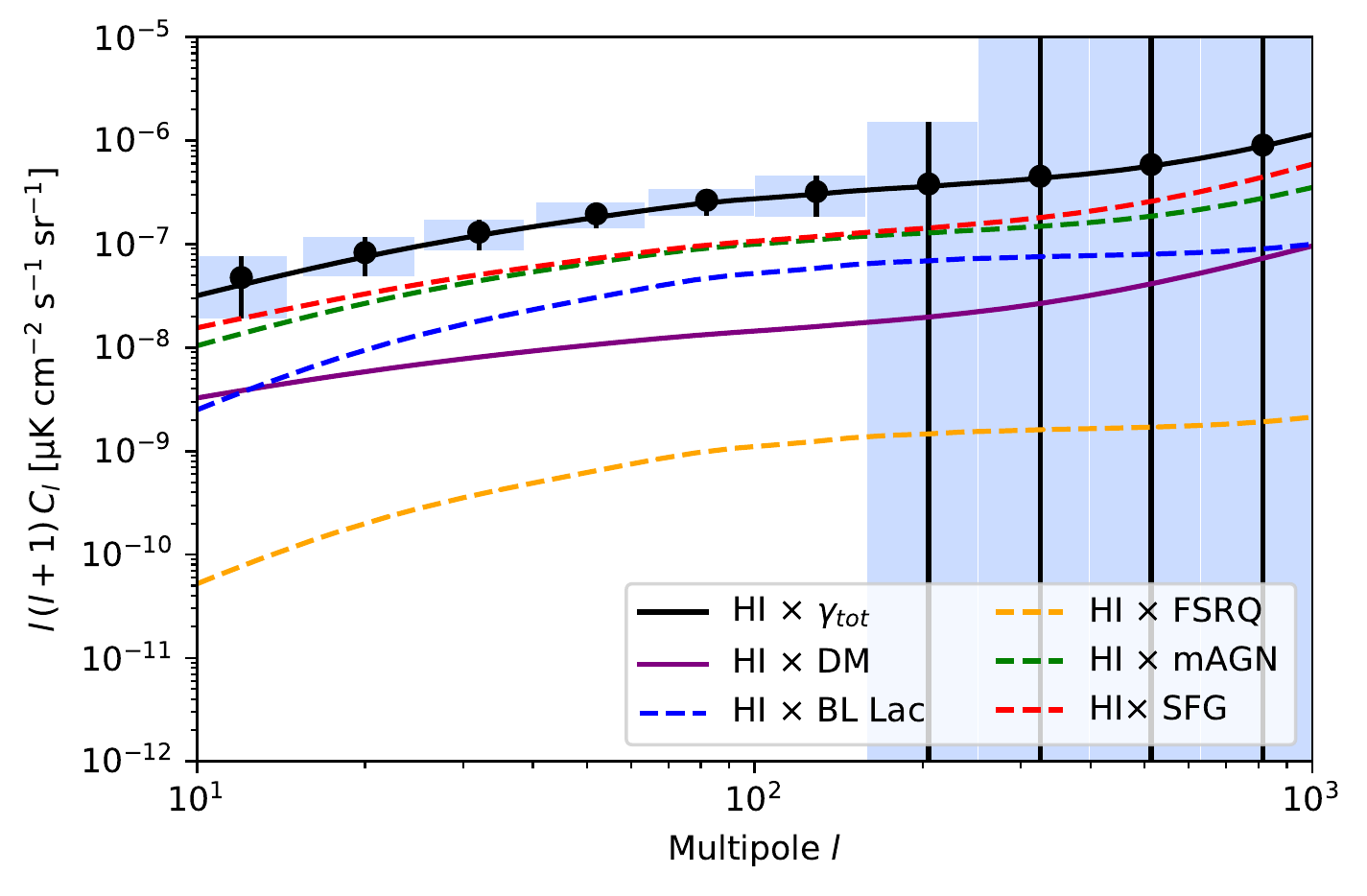}
\caption{\label{fig:Cl} Cross-correlation angular power spectrum between the 21-cm brightness temperature of neutral hydrogen and the unresolved gamma emission. Dashed lines denote the astrophysical components, purple solid line refers to DM annihilation with mass $m_{\chi}$ = 100 GeV and annihilation cross-section $\langle \sigma v \rangle = 3 \times 10^{-26}\,\mathrm{cm^3\,s^{-1}}$. The solid black line is the total signal. In the right panel, error bars are derived for the the combination {\it Fermi}-LAT $\times$ MeerKAT in the case of the single-dish set-up and lower-redshift band ($z<0.58$). In the left panel, we consider the combination {\it Fermi}-LAT $\times$ SKA and the lower-redshift band ($z<0.5$). Note that as a result of the improved specifications of the radio telescope, the error bars are smaller as compared to those of MeerKAT.}
\end{figure*}

\begin{figure}[t]
\begin{center}
\includegraphics[width=0.45\textwidth]{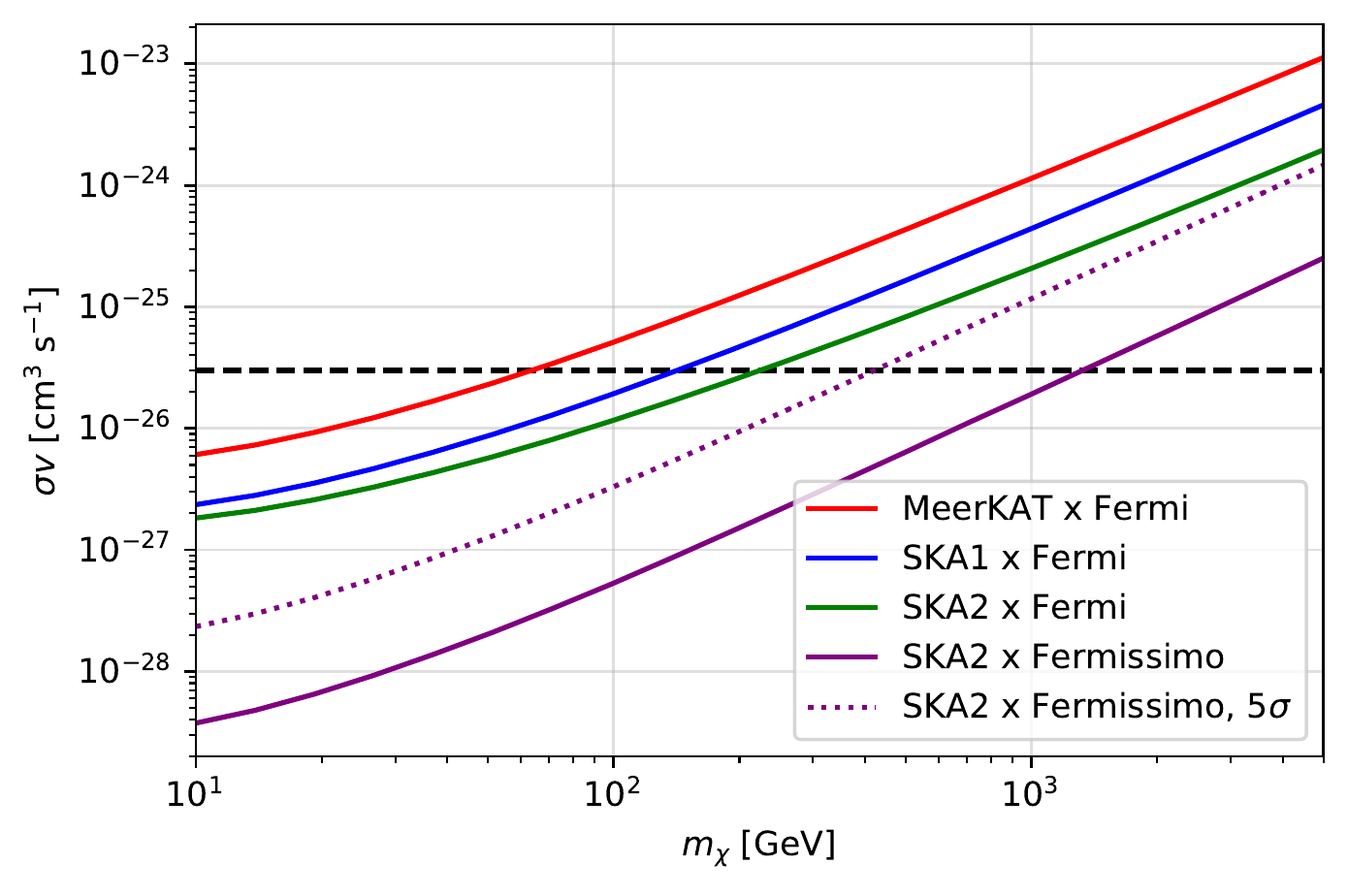}
\end{center}
\caption{\label{fig:bound} Expected upper bounds in the DM parameter space which we can obtained with the combination {\it Fermi}-LAT $\times$ MeerKAT (red), {\it Fermi}-LAT $\times$ SKA Phase 1 (blue), {\it Fermi}-LAT $\times$ SKA Phase 2 (green) at 2$\sigma$.  The solid purple line refers to the bounds attainable with a next-generation gamma detector {\it Fermissimo} at 95\% C.L. while the dotted purple line refers to the 5$\sigma$ detection bounds. The dashed black curve remarks the thermal annihilation cross-section $\langle \sigma v \rangle = 3 \times 10^{-26}\,\mathrm{cm^3\,s^{-1}}$.}
\end{figure}

The cross-correlation signal has been summed over 12 gamma-ray energy bins from 0.5 GeV to 1 TeV and integrated in two redshift bands, corresponding to low redshift ($z<0.58$ for MeerKAT and $z<0.5$ for SKA) and high redshift ($0.4<z<1.45$ for MeerKAT and $0.35<z<3$ for SKA). Moreover, two set-ups have been considered for the radio telescopes: single-dish and the combination single-dish+intereferometer [1]. Fig. \ref{fig:Cl} shows the APS of the cross-correlation between the 21-cm brightness temperature and the unresolved gamma emission at low redshift for the combination of the space telescope {\it Fermi}-LAT with the single-dish configuration of MeerKAT and SKA, respectively. The dashed curves refer to the astrophysical contributions, the purple solid line shows the signal originated by annihilation events of DM particles with thermal cross-section $\langle \sigma v \rangle = 3 \times 10^{-26}\,\mathrm{cm^3\,s^{-1}}$ and mass $m_{\chi}$ = 100 GeV for the $b \bar b$ channel, the solid black line denotes the total signal due to the sum of all the contributions. As expected, the astrophysical sources are the leading contribution to the total signal, however the DM component at low redshift is only a factor of 3 to 5 lower. The improved specifications of SKA as compared to its precursor MeerKAT considerably reduce the size of the error bars. For an extensive discussion on the error estimation as well as for the cross-correlation signal in the case of the higher-redshift band and combined configuration, please refer to Ref. [1].
To figure out the potential detectability of a cross-correlation signal purely of astrophysical origin (in order to be independent on any hypothesis on the DM particles), it is convenient to employ the signal-to-noise ratio (SNR) which can be written as 

\begin{equation}
{\rm SNR}^2= \sum_{l,a} \left(\frac{C_{l,a}^{HI\gamma_\star}}{\Delta C_{l,a}^{HI\gamma_\star}}, \right)^2
\end{equation}

where the sum is performed over the multipoles $l$ from 10 to 1000 and over the 12 gamma-ray energy bins $a$ within each redshift band under consideration (corresponding to low and high redshift). The combination {\it Fermi}-LAT $\times$ MeerKAT has the potential to reach a SNR of 3.6 at low redshift and 3.7 at high redshift for both configurations, while with SKA the SNR is predicted to reach 5.7 at low redshift for both set-ups. The combined configuration single-dish+interferometer becomes particularly favourable in the case of a radio telescope with enhanced capabilities, allowing to achieve a SNR of 8.2 at high redshift. Consequently, we can safely claim that a cross-correlation signal is likely detectable.
Thereafter, we can scrutinize whether a DM signal is identifiable above the leading astrophysical contribution by adopting as a convenient statistics

\begin{equation}
\Delta \chi^2 = \sum_{l,a} \left(\frac{C_{l,a}^{HI\gamma_{\star +DM}}}{\Delta C_{l,a}^{HI\gamma_{\rm \rm \star+DM}}}\right)^2 - \sum_{l,a} \left(\frac{C_{l,a}^{HI\gamma_\star}}{\Delta C_{l,a}^{HI\gamma_\star}}\right)^2,
\end{equation}
which corresponds to discriminating the null hypothesis (gamma emission only of astrophysical origin $\gamma_{\star}$) against the alternative hypothesis (both DM and the astrophysical sources $\gamma_{\star + DM}$ contribute to the total signal). This quantity is distributed as a $\chi^2$ with only one degree of freedom corresponding to the annihilation cross-section and can be exploited to derive the bounds at 95\% C.L. (i.e. 2$\sigma$) by requiring $\Delta \chi^2$ = 4. Fig. \ref{fig:bound} shows the upper bounds attainable with the combined set-up in the lower-redshift band for the statistics of {\it Fermi}-LAT combined with MeerKAT (red line), SKA Phase 1 (blue), SKA Phase 2 (green). Already with MeerKAT the attainable bounds are competitive as compared to the present panorama. However, SKA could do even better, potentially allowing to probe a thermal DM candidate up to masses of few hundreds GeV, respectively. The solid purple line shows the bounds that could be derived with an ungraded radio telescope and an hypothethical next-generation gamma-ray space telescope characterized by a better angular resolution and a larger exposure \cite{Pinetti2019}: it would have the potentiality to probe the thermal cross-section of a DM particle for masses up to the TeV scale at 2$\sigma$ and up to 400 GeV at 5$\sigma$ (dotted purple line).   

\section{Conclusions}
In this work we investigated the opportunities offered by the cross-correlation angular power spectrum between the 21-cm brightness temperature of neutral hydrogen and the intensity of the unresolved gamma-ray emission by considering as benchmark experiments {\it Fermi}-LAT as concerns gamma rays, MeerKAT and SKA as regards intensity mapping of neutral hydrogen. It is noteworthy that the combination {\it Fermi}-LAT $\times$ SKA has the capability to identify a signal of astrophysical origin with a signal-to-noise ratio above 5$\sigma$. The cross-correlation signal can be exploited to derive the bounds in the DM parameter space. The sensitivity of {\it Fermi}-LAT combined with SKA could probe a thermal DM particle up to masses of 130 GeV, while with an ungraded radio telescope it would be possible to reach 200 GeV at 95\% C.L. and with a next-generation gamma-ray telescope we would be able to extend the attainable bounds up to the TeV scale at 2$\sigma$ and up to 400 GeV at the 5$\sigma$ detection level.    


\smallskip


\section{References}
\begin{thebibliography}{9}
\bibitem{Pinetti2019} Pinetti E, Camera S, Fornengo N and Regis M 2019 arXiv: 1911.04989
\bibitem{Camera2013} Camera S, Fornasa M, Fornengo N and Regis M 2013 {\it Astroph. J.} {\bf 771} L5 arXiv: 1212.5018
\bibitem{Santos} Santos M et al {\it Proc. Advancing Astrophysics with the Square
                        Kilometre Array} June 9-13 2014 Giardini Naxos

 

\end{thebibliography}
\end{document}